# Rainbow Scattering from Graphene

Carolin Frank[1,2*], Kevin Vomschee[1], Radek Holeňák[3], Yossarian Liebsch[2], Marika Schleberger[2], and Daniel Primetzhofer[1,3]

[1]*Materials Physics, Department of Physics and Astronomy, Uppsala University, Uppsala, Sweden*
[2]*Faculty of Physics and CENIDE, University of Duisburg-Essen, Duisburg, Germany*
[3]*Tandem Laboratory, Uppsala University, Uppsala, Sweden*

*Corresponding author: carolin.frank@physics.uu.se

## Abstract

We report the experimental observation of atomic rainbow scattering of 40 keV $Xe^+$ ions transmitted through self-supporting single-layer graphene using time-of-flight medium energy ion scattering. Supported by molecular dynamics and binary collision approximation simulations, we show that the rainbow pattern of graphene consists of a small hexagonal inner rainbow, arising from projectiles with characteristic trajectories interacting with multiple carbon atoms, and a larger circular outer rainbow, arising from close binary collisions between projectiles and individual carbon atoms.

## Article Text

The rainbow effect is ubiquitous in various scattering theories of both waves and particles, appearing in contexts ranging from the familiar meteorological phenomenon to nuclear and crystal scattering phenomena. The atomic rainbow effect, first described for semiclassical particle scattering by Ford and Wheeler in 1959, arises when projectiles with different impact parameters are scattered into the same characteristic angle due to an extremum of the deflection function [1]. This extremum leads to a caustic (an infinite density of classical trajectories) in angular space, resulting in singularities in the classical differential cross section. Consequently, scattering patterns provide information on the projectile-target interaction potential.

In previous years, rainbow scattering in specular reflection of neutral atoms from surfaces was observed in combination with diffraction patterns, providing insights into projectile-surface interaction [2], [3]. More recent studies of projectile-target interaction focus on diffraction patterns arising from transmission of projectiles through thin films. Discovered by Novoselov and Geim in 2004 [4], graphene is a highly promising material for applications in electronics due to its exceptional electrical, mechanical and thermal properties, and is ideally suited for such transmission experiments due to its monolayer structure and high crystalline quality. Single-layer graphene (SLG) has a hexagonal structure of carbon atoms, corresponding to a single scattering regime in transmission experiments. Diffraction of fast atoms through SLG, just experimentally realized by Kanitz et al. and Guichard et al., revealed diffraction patterns of hydrogen and helium atoms [5], [6], highlighting the unique phenomena accessible in transmission experiments.

Rainbow scattering in graphene is expected to provide insights not only into ion-atom interaction potentials, but also into the sample's composition and structure. Several attempts to simulate rainbow scattering in graphene have been made. Recently, simulations for protons transmitted through SLG have been reported [7], [8], [9], [10]. Brand et al. simulated the characteristic rainbow scattering pattern of 5 keV protons transmitted through SLG based on classical trajectories and using the ZBL potential [7]. The simulated results show that graphene's rainbow pattern consists of a hexagonal pattern arising from projectiles with characteristic trajectories interacting with multiple carbon atoms. However, the experimental observation of rainbow scattering in graphene is challenging due to two essential requirements [10], [11]:

- An unsaturated detector of a very high angular resolution is required to observe a detailed rainbow pattern.
- A single-layer self-supporting graphene of a very high cleanliness is required. Even smallest amounts of contamination – typically introduced in the transfer process – will inevitably enhance multiple scattering of projectiles. Rainbow scattering is sensitive to both orientation and defects of the single-layer graphene as well as to the lattice site of adsorbates. A damaged graphene sheet will result in a blurred scattering pattern due to enhanced multiple scattering.

In this Letter, we demonstrate the first unambiguous experimental observation of rainbow scattering patterns in ion transmission through single-layer graphene. Based on spatial 2D scattering distributions of scattered projectiles, we show that the rainbow scattering pattern of 40 keV $Xe^+$ ions transmitted through SLG consists of two parts: An inner rainbow with a hexagonal pattern, created by projectiles passing through characteristic regions in a graphene honeycomb experiencing small angle deflection by several carbon atoms simultaneously, and an outer rainbow, which stems from projectiles experiencing close collisions with individual carbon atoms. To support our findings, we present comparative simulations based on Molecular Dynamics (MD) for both the Ziegler-Biersack-Littmark (ZBL) [12] and the Nordlund-Lehtola-Hobler (NLH) [13] potential, as well as based on the Binary Collision Approximation (BCA).

As a target system we used single-layer flattened graphene freely suspended on a transmission electron microscopy (TEM) grid. Graphene on copper was purchased from Graphenea and transferred to a gold TEM grid with a 10-15 nm thick amorphous carbon support foil (Quantifoil R 1.2/1.3 Au grid) following the polymethyl methacrylate (PMMA)-free flattening transfer process described in detail in reference [14]. In the target system, regions of self-supporting graphene as well as regions of graphene supported by Quantifoil are present. The degree of coverage is approximately 90%. As discussed below, ions are transmitted exclusively through graphene.

Transmission experiments were performed at the Time-of-Flight Medium Energy Ion Scattering (ToF-MEIS) setup at Uppsala University [15], [16]. A detailed description of sample pretreatment and technical details is included in the Supplementary Information [17].

For the ToF-MEIS measurements, a pulsed beam of singly charged $^{129}Xe^+$ ions with a kinetic energy of 40 keV is transmitted through the target with the monolayer graphene facing the ion beam. Transmitted particles are detected by a circular position-sensitive microchannel plate (MCP, DLD120 from RoentDek [19]), which is positioned 290 mm behind the sample. For this study, the evaluated detector region was limited to a circle with a radius of 40 mm around the position of the direct beam, covering scattering angles of 7.85°. The recorded time-of-flight spectrum for 40 keV $^{129}Xe^+$ ions transmitted through SLG on a gold TEM grid with Quantifoil is shown in Figure 1, including a schematic of the experiment.

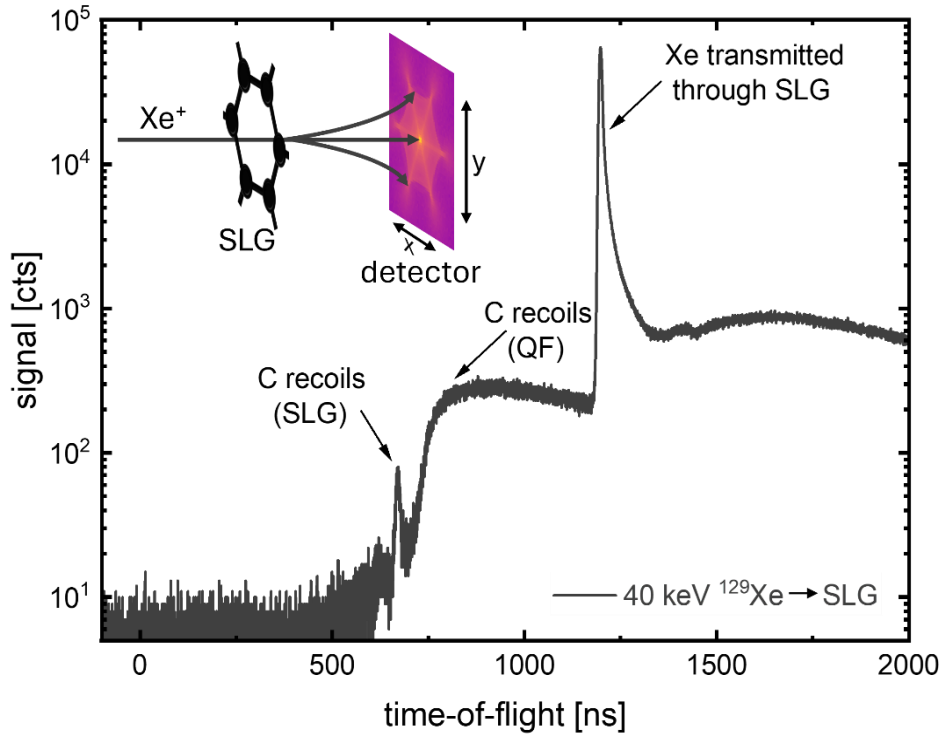

Figure 1: Time-of-flight spectrum of 40 keV $Xe^+$ ions transmitted through SLG, including a schematic of the experiment. Dominant peaks are attributed to C recoils from the graphene layer, to C recoils from the Quantifoil and to Xe ions transmitted through SLG.

The peak at 672 ns can be attributed to carbon recoils from the target. A detailed description of elastic recoil detection analysis in transmission time-of-flight ion scattering is given in reference [20]. At 40 keV, no Xe ions are transmitted through the Quantifoil. The dominant peak at a flight time of 1198 ns corresponds to Xe ions transmitted through the single-layer graphene, including the direct beam passing through pin holes and damaged areas of the sample. Since virtually no H recoils – expected at 599 ns – are present in Figure 1, we consider the sample to have a high cleanliness throughout the measurement. To evaluate scattering of Xe ions after transmission through SLG, the spatial distribution of transmitted projectiles on the detector was analyzed for the time-of-flight range corresponding to the dominant graphene peak.

Experimental scattering distributions were compared to scattering distributions simulated using the Molecular Dynamics (MD) Kalypso code [21]. MD simulations explicitly resolve the correlated, time-dependent many-body dynamics. Molecular dynamics simulations were performed using the widely used Ziegler-Biersack-Littmark (ZBL) potential [12] and the recently developed Nordlund-Lehtola-Hobler (NLH) potential, which is based on density functional theory calculations and demonstrated good agreement with experimental results in the energy range from 15 keV to 200 keV [13]:

- ZBL potential: $V(r) = \frac{Z_i Z_j e^2}{4\pi\varepsilon_0 r_{ij}} \sum_{k=1}^{4} c_k e^{\frac{-b_k r_{ij}}{a}}$
- NLH potential: $V(r) = \frac{Z_i Z_j e^2}{4\pi\varepsilon_0 r_{ij}} \sum_{k=1}^{3} c_k e^{-b_k r_{ij}}$

Here, $Z_{i,j}$ denote the atomic numbers of the collision partners, $e$ in the prefactor of the summation is the elementary charge, $\varepsilon_0$ is the dielectric constant, $r_{ij}$ is the distance between collision partners and $a$ is the screening length. For the NLH potential, we used the coefficients $b_k$ and $c_k$ provided in the supplementary material of reference [13] for Xe and C. Note that the $b_k$ parameters have different dimensions in the two potentials.

Simulations were performed at a thermal displacement of 300 K and 500 K for both ZBL and NLH potential to investigate the impact of the thermal displacement on rainbow scattering. Since no differences between MD-simulated scattering distributions for 300 K and 500 K were observed, we show results for a thermal displacement of 300 K in the present Letter and include simulation results for 500 K in the Supplementary Information [17].

Both experimental and MD simulated scattering distributions were compared to simulated ones using the Binary Collision Approximation (BCA) code SIIMPL developed by Janson [22]. In contrast to MD simulations, BCA simulations reduce the interaction potential to independent sequential binary collisions and do not consider collective effects. Here, BCA simulations were performed using the Magic Formula developed by Biersack et al. [12] and assuming the ZBL potential. Simulation parameters for both MD and BCA simulation are included in the Supplementary Information [17].

In Figure 2a), we present the measured 2D distribution of 40 keV Xe[+] ions transmitted through a self-supporting SLG onto the detector plate. For comparison with Figure 2a), the MD-simulated 2D distribution of 40 keV Xe[+] ions transmitted through self-supporting SLG is shown in Figure 2b), using a ZBL potential at random thermal displacements as expected at 300 K. The integrated signal, expressed in counts, is color-coded as a function of the detector's x and y coordinates. The corresponding scattering angle is indicated on the upper and right axes.

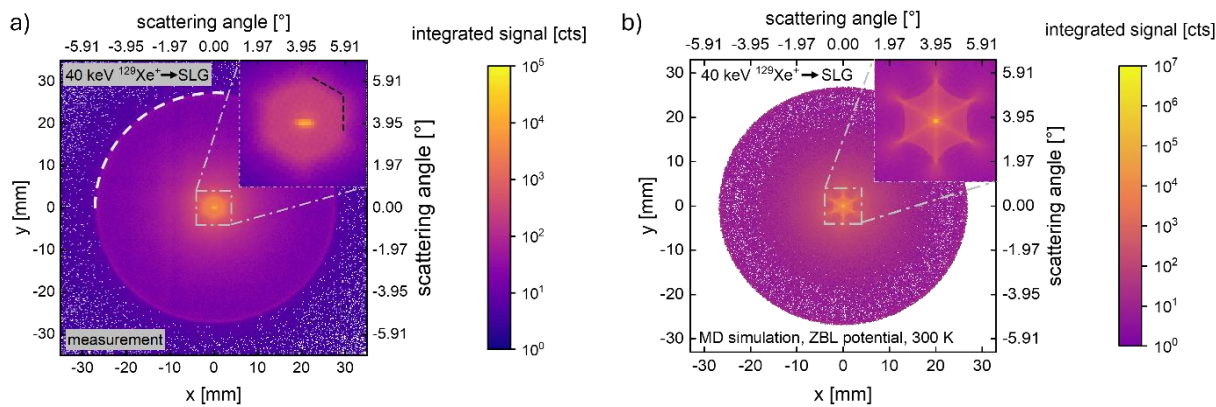

Figure 2: a) Measured 2D distribution of 40 keV Xe$^+$ ions transmitted through SLG. As a visual guide, the circular shape of the outer rainbow and the hexagonal shape of the inner rainbow, enlarged in the inset, are indicated with dashed lines. b) MD simulated 2D distribution of 40 keV Xe$^+$ ions transmitted through SLG using the ZBL potential at a thermal displacement of 300 K. The color scale of the inset in a) and b) is adjusted to enhance the visibility of the features.

Both Figure 2a) and 2b) exhibit a distinct and extremely sharp edge at radius of 27 mm, indicated with a dashed line as visual guide in Figure 2a). This local maximum, referred to as the *outer rainbow* in the following, corresponds to the maximum deflection of Xe projectiles in a binary collision by a single C atom. The observed maximum deflection angle is 5.32°; the classical mechanics limit is 5.34°.

Moreover, both measured and MD-simulated 2D distribution exhibit a hexagonal pattern, referred to as the *inner rainbow* in the following. As visual guide, the shape of the hexagon is indicated with dashed lines in the inset in Figure 2a). The measured inner rainbow has an inradius of 2.4 mm; the MD-simulated inner rainbow has an inradius of 1.5 mm. The inner rainbow stems from projectiles with trajectories passing through high-symmetry regions in a graphene honeycomb experiencing small angle deflection by several carbon atoms simultaneously. The fact that the hexagon is also visible in the measured spatial 2D distribution indicates that the measured graphene region consists mainly of an atomically clean single domain, consistent with ongoing experiments. A reason for the slightly blurred edge of the hexagon might be the presence of small crystal domains in different orientations, as graphene grown via chemical vapour deposition is known to exhibit domains of different orientation.

For both, measurement and MD simulation, the global maximum of the integrated signal is found at the center of the scattering distribution (0 mm| 0 mm). For the measurement, this corresponds to the position of the direct beam transmitted through pinholes in the graphene.

In Figure 3, we present the derived angular scattering distributions of 40 keV Xe$^+$ ions transmitted through self-supporting SLG obtained from a typical measurement (grey), together with angular scattering distributions obtained from MD-simulations using the ZBL potential for 300 K (blue), the NLH potential for 300 K (green), as well as BCA-simulations (purple). For better comparability, the integrated signal height of both measured and BCA-simulated curves is adjusted to the maximum integrated signal of the MD-simulated curves at 5.27°. The uncorrected measurement data as a function of the scattering angle is shown in the Supplementary Information [17]. In the inset of Figure 3, we show the scattering distributions in the range from 0° to 0.6°.

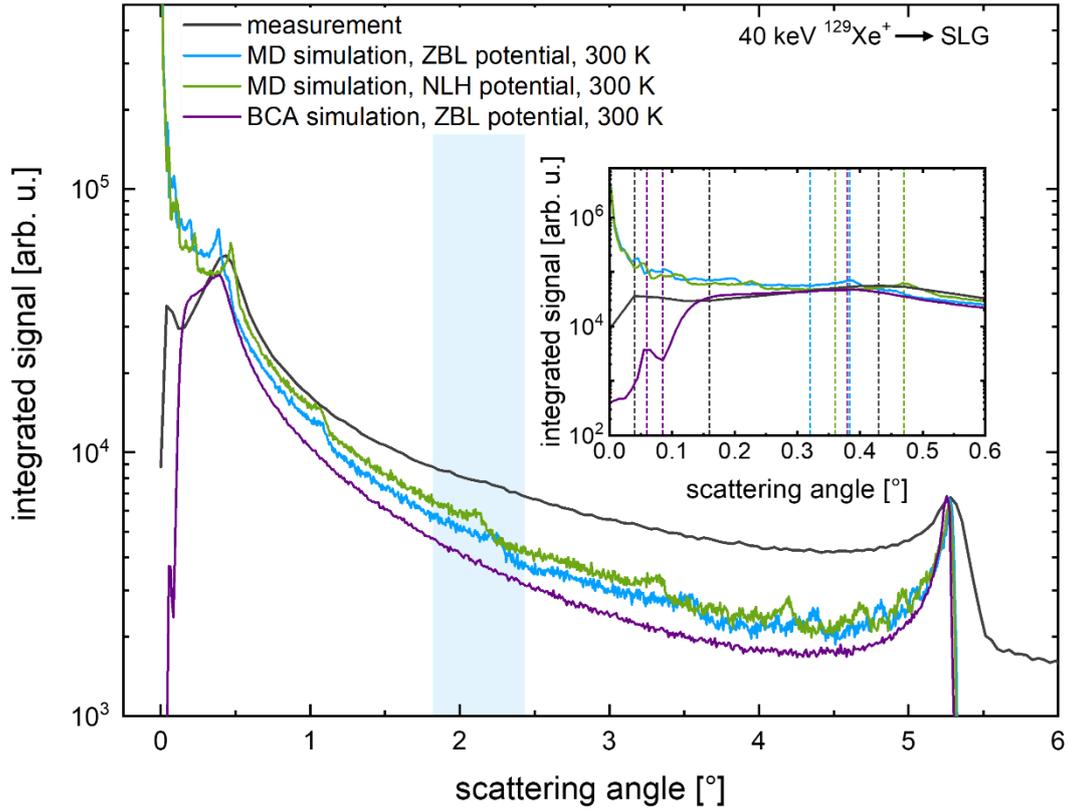

Figure 3: Integrated signal as a function of the scattering angle as measured (grey), BCA-simulated (purple) and MD-simulated using the ZBL potential (blue) and the NLH potential (green). For better comparability, the integrated signal height of both measured and BCA-simulated curves is adjusted to the maximum integrated signal of the MD-simulated curves at 5.27°.

The angular scattering distributions shown in Figure 3 differ from each other in terms of scattering angles at which local maxima of the integrated signal occur, the number of maxima, and the shape of the peaks.

In the range from 4.5° to 5.5°, all curves exhibit a peak at similar scattering angles, which corresponds to the outer rainbow (measured: 5.28°, BCA-simulated: 5.25°, MD-simulated: 5.28° for ZBL potential, 5.27° for NLH potential). While all simulated curves show a sharp edge around 5.3°, the edge of the measured peak is smeared out, which is explained predominantly by the divergence of the $Xe^+$ ion beam. Based on the comparison of measurement and simulations, we conclude that the outer rainbow is independent of the used interaction potential since scattering kinematics is unaffected by conservative potentials.

In the range from 0.68° to 4.5°, the integrated signal decreases for all curves. In this range, the measured integrated signal is larger than the simulated signals. In comparison to the BCA-simulated curve, the MD-simulated curves show maxima which depend on the used potential (e.g. at 1.08° for ZBL potential and at 1.05° for NLH potential). Around a scattering angle of 2°, a weak feature in the form of locally enhanced intensity is also visible in the measured angular distribution. The respective region is highlighted in Figure 3). Analysis of standardized residuals revealed local deviations of 3-4 standard deviations for three consecutive data points from the fitted function. We attribute these additional peaks in the MD simulations and in the measurement to *secondary rainbows* which occur for projectiles experiencing the saddle point in the potential of two carbon atoms in the middle of the honeycomb's edge. Due to the different dependence on the distance $r_{ij}$, the respective scattering angles differ from each other for ZBL and NLH potential in the MD simulation. Secondary rainbows are also visible in the MD-simulated spatial distribution in Figure 2b) in the form of additional circles with local maxima at radii of 6 mm, 12 mm and 18.6 mm. In the BCA-simulated angular scattering distribution, no additional peaks are observed because only the potential of a single carbon atom is considered.

For small scattering angles starting from 0°, both the measured and BCA-simulated scattering distributions rise sharply, reaching a maximum in the integrated signal (measurement: 0.04°, BCA simulation: 0.06°), followed by a local minimum (measurement: 0.16°, BCA simulation: 0.08°). In contrast to that, the MD-simulated integrated signal decreases from a maximum at 0°, reaching a minimum at 0.32° for the ZBL potential and at 0.36° for the NLH potential. All curves show a dominant peak in the range from 0.3° to 0.5°, corresponding to the inner rainbow. In the measurement, the inner rainbow is found at a scattering angle of 0.43, while in the BCA simulation it shifted to 0.35°. In contrast to the outer rainbow, the inner rainbow depends on the applied interaction potential (ZBL: 0.38°, NLH: 0.47°) due to the different dependence of the respective potential on the distance between projectile and C atom. In the BCA simulation, the inner rainbow arises because the set of impact parameters is limited to the hexagonal Wigner-Seitz cell of the graphene crystal. Hence, in the BCA simulation, the integrated signal decreases towards zero at a scattering angle of 0°. In the MD simulations, these effects are supplemented by the superposition of the potentials of the surrounding carbon atoms. Due to the superposition of the potentials and the resulting rather flattened potential landscape between neighboring carbon atoms, the MD simulations show a strong increase towards a maximum value of the integrated signal around a scattering angle of 0°.

Radially symmetric potentials exhibit several symmetry points when superimposed onto each other in graphene, resulting in 0° deflection being likely, as observed in the MD simulations. These symmetry points associated with 0° scattering scenarios – such as central impact on a C atom, passage through the center of a hexagon, or through the midpoint between two C atoms – are illustrated schematically in the Supplementary Information [17]. However, the measured angular distribution shows a decreasing signal towards 0°, indicating that 0° scattering is rare despite being geometrically possible. This observation can, on the one hand, be attributed to the actual potential being non-radially symmetric – e.g., due to chemical bonds – in particular at comparably large interaction distances relevant for deflections near 0°. On the other hand, the scattering regime near 0° is highly sensitive to electronic processes: charge exchange between Xe and SLG alters the effective interaction potential along the projectile's trajectory [23], [24, 25], which thus varies for each projectile and is particularly important for two-dimensional target systems in the single scattering regime. Accounting for associated momentum transfers to the electrons, leading to small angle deflections, further reduces the probability of 0° scattering. We therefore conclude that averaged, radially symmetric potentials, as assumed in the MD simulations, are insufficient to accurately describe the here investigated projectile-target interaction.

We have observed the rainbow scattering pattern of 40 keV Xe$^+$ ions transmitted through self-supporting single-layer graphene. Supported by MD and BCA simulations, we find that the measured rainbow pattern consists of two parts: Projectiles passing through characteristic regions in a graphene honeycomb experiencing small angle deflection by several carbon atoms simultaneously scatter in a small hexagonal pattern due to the non-radial symmetry, whereas projectiles experiencing close collisions with carbon atoms scatter in a larger, circular pattern. The outer rainbow at 5.28° is determined solely by kinematics, while the inner rainbow, observed at 0.43°, represents the combined influence of multiple carbon atoms and the associated electronic system: the measurements in the near-0° deflection regime – highly sensitive to electronic processes – demonstrate that averaged, radially symmetric potentials are insufficient to describe projectile-target interactions, as charge exchange processes and resulting small variations of the actual potential play a decisive role in the interaction.

With this study, we open the door to investigate interatomic potentials and their dynamics in solids in unprecedented detail via transmission experiments. Our work can easily be extended to include more complex interaction scenarios such as experiments with multiply charged ions, molecular beams or 2D materials beyond graphene, e.g. transition metal dichalcogenides. These opportunities will provide insights into electronic structures and screening effects specific to 2D materials – and exceptional access to semiclassical dynamics.

## Acknowledgements

Accelerator operation is supported by the Swedish Research Council VR-RFI (Contract No. 2023-00155). C. F. and M. S. acknowledge financial support from the DFG (CRC 1242 "Non-Equilibrium Dynamics of Condensed Matter in the Time Domain", project number 278162697, and "2D Mature", project number 461605777) and from the Faculty of Physics.